\documentclass[a4paper]{jpconf}
\usepackage{graphicx}
\begin{document}
\title{Looking forward: exclusive dilepton production with a leading proton}
\author{Michele Gallinaro$^{1,2}$}
\address{$^1$on behalf of the CMS and TOTEM collaborations}      
\address{$^2$Laborat\'orio de Instrumenta\c{c}\~ao e F\'isica Experimental de Part\'iculas, LIP Lisbon, Portugal}

\ead{michgall@cern.ch}

\begin{abstract}
Exclusive dilepton production occurs with high cross section in gamma-mediated processes at the LHC. The pure QED process $\gamma\gamma\rightarrow\ell^+\ell^-$ provides
the conditions to study particle production with masses at the electroweak scale.
By tagging the leading proton from the hard interaction, the Precision Proton Spectrometer (PPS) provides an increased sensitivity to selecting exclusive processes. 
PPS is a detector system to add tracking and timing information at approximately 210~m 
from the interaction point around the CMS detector. 
It is designed to operate at high luminosity with up to 50 interactions per 25~ns bunch crossing to perform measurements of 
e.g. the quartic gauge couplings and search for rare exclusive processes. Since 2016, PPS has been taking data in normal high-luminosity proton-proton LHC collisions.
Exclusive dilepton production with proton tagging, the first results obtained with PPS, and the status of the ongoing program are discussed.
\end{abstract}

\section{Introduction}

Recently, there has been a renewed interest in studies of central exclusive production (CEP) processes in high-energy proton-proton collisions. 
CEP provides a unique method to access a variety of physics topics, such as new physics via anomalous production of W and Z boson pairs, high transverse momentum 
($p_T$) jet production, and possibly the production of new resonances. 
These studies can be carried out in particularly clean experimental conditions thanks to the absence of proton remnants.

CEP of an object X may occur in the process $pp\rightarrow p + X + p$, where "+" indicates the ``rapidity gaps" adjacent to the state $X$. 
Rapidity gaps are regions without primary particle production.
In the high mass region with both protons detected, among some of the most relevant final states are $X = e^+e^-,\mu^+\mu^-,\tau^+\tau^-$ and $W^+W^-$. 
In CEP processes, the mass of the state $X$ can be reconstructed from the fractional momentum loss $\xi_1$ and $\xi_2$ of the scattered protons
by using the expression $M_X=\sqrt{\xi_1\cdot \xi_2\cdot s}$.
The $M_X$ reach at the LHC is significantly larger than at previous colliders because of the larger $\sqrt{s}$.
The scattered protons can be observed mainly thanks to their momentum loss, due to the horizontal deviation from the beam trajectory. 
The acceptance in $\xi$ depends on the distance from the IP and on how close to the beam the proton detectors can be moved.
For the first time, proton-proton collisions at the LHC provide the conditions to study particle production with masses at the electroweak scale through photon-photon fusion.
At $\sqrt{s}=13$~TeV and in normal high-luminosity conditions, values of $M_X$ above 300~GeV can be probed.
CEP processes at these masses have small cross sections, typically of the order of a few fb, and thus can be studied in normal high-luminosity fills.

\section{The Precision Proton Spectrometer}

The Precision Proton Spectrometer (PPS)~\cite{ctppstdr} 
%is a joint project of the CMS and TOTEM collaborations 
allows
to measure the surviving scattered protons during standard running conditions in regular "high-luminosity" fills. 
It adds precision tracking and timing detectors in the very forward region 
on both sides of the CMS detector at about 210 meters from the interaction region to study CEP in proton-proton collisions.
The PPS detector operated for the first time in 2016 and has been collecting data in normal LHC high-luminosity conditions since then.
It will provide an opportunity to explore a new window into high-$Q^2$ physics processes with a unique sensitivity to physics beyond the standard model (BSM).

The PPS detector consists of a silicon tracking system to measure the position and direction of the protons, and a set of timing counters to measure their arrival time.
This allows the reconstruction of the mass and momentum as well as of the $z$ coordinate of the primary vertex of the centrally produced system.
The protons that have lost a small fraction of their momentum are bent outside the beam profile by the LHC magnets between the
interaction point 
and the detector stations, and their trajectories can be measured in the PPS detector.
The detector covers an area transverse to the beam of about 4~cm$^2$ on each arm. It uses a total of 144 pixel readout chips and about 100 timing readout channels.
The detectors are housed in Roman Pots (RP) that allow moving the sensitive detectors into the vacuum of the LHC, in such a way that they get to approximately 1-2~mm from the beam. The tracking detectors are placed in two stations located 10~m apart, on either side of the collision point. Six planes of silicon pixels in each station can detect the track of the scattered protons to provide the track direction information. Detectors are tilted with respect to the beam axis in the $x-z$ plane in order to recover the full efficiency~\cite{ctppstdr}. 
3D sensors are used for their improved resistance to radiation. Readout is based on that developed for the Phase-1 upgrade of the CMS silicon pixel detectors.
A spatial resolution of 17~$\mu$m was measured in data, consistent with test beam results.
Due to the non-uniform irradiation, few of the pixel sensors closer to the beam suffer an increased radiation damage.
In order to recover the full efficiency, the detectors can be moved vertically.

One RP station on each side is instrumented with timing detectors.
Located in both arms, they measure the difference of arrival time of the two outgoing protons. A time resolution of 10~ps corresponds to a vertex resolution of 2.1~mm, thus allowing pileup rejection.
The timing detector technology is based on single-crystal CVD diamond sensors readout by custom-designed dedicated electronics. The geometry is optimized to minimize the detector occupancy, with smaller sensors closer to the beam where the occupancy is higher, and larger sensors away from the beam where the occupancy is lower. 
A time resolution of 80~ps was measured in a test beam for a single plane~\cite{Antchev:2017pjj}.
After further R\&D studies, a time resolution of 50~ps was measured for a double-sided diamond detector~\cite{Berretti:2016sfj}, an improvement of 1.6 over the single-sided sensor.
In a RP station on each side, one plane instrumented with one double-sided and three planes of single-sided diamond detectors are currently used in the 2018 data-taking.

Regular detector operation in high-luminosity conditions was required in order to successfully carry out the physics program. 
However, this presented several challenges due to the harsh conditions of radiation and beam backgrounds.
The ability to operate the detectors closer to the beam center enhances the sensitivity to lower invariant mass systems. 
In the current conditions, the detectors are operating at a distance of 1-2~mm from the beam, which corresponds to $\approx 15\sigma$ of the beam dispersion.
However, a close approach is limited by the interference with the beam,
as well as the detector damage caused by the high radiation levels.
Studies based on radiation monitors placed on the 220~m RP stations yield a total dose of about 100~Gray and a fluence of the order 
of $10^{12}~n_{\rm eq}$cm$^{-2}$ per 100~fb$^{-1}$ of integrated luminosity. 
These are values at the location of front-end electronics in the RPs. 
A proton flux of up to $5\times10^{15}$cm$^{-2}$ is expected on the sensors themselves.
Another challenge comes from the large background expected in the presence of multiple interactions (pileup) per bunch crossing. 
The number of pileup interactions per bunch crossing was approximately $\mu$=20$\div$30 during 2016-2017, and $\mu=40$ during 2018. It may reach 50 or 100 in the next few years.
The dominant background results from inelastic events overlapping with two protons from single diffraction events occurring in the same bunch crossing. 
This pileup background can be reduced by using proton timing information to determine the z-position of the primary vertex.
In the future, the expected increased pileup and large radiation doses represent some of the difficult challenges that need to be addressed with a focused upgrade program.
Upgrades to the tracking and timing detector systems and associated electronics are foreseen.
These are needed in order to maintain the physics goals and be able to efficiently collect the large samples of data necessary to explore the rare processes.

\section{Exclusive dilepton production and prospects}

Proton-proton collisions at the LHC provide for the first time the conditions to study particle production with masses at the electroweak scale via photon-photon fusion.
The LHC is known for colliding protons but is also a photon-photon collider with a unique energy range of $\sqrt{s_{\gamma\gamma}}$ up to approximately 1~TeV, a region so far unexplored.
The exclusive two-photon production of lepton pairs can be calculated in the framework of QED with great precision. Exclusive dilepton events have a clean
signature that helps discriminate them from background: there are only two identified muons or electrons
without any other activity in the central detectors, and the leptons are back-to-back in azimuthal angle.
After the collision, the two outgoing protons remain intact and recoil as a result of the photon-photon interaction.
It is possible to determine whether the photon interactions took place by identifying these deflected protons in the PPS detector, thus effectively treating the LHC as a photon collider and adding a new probe for exploring fundamental physics.
The detection of the two final state protons, scattered at almost zero-degrees, in the very forward near-beam PPS detectors, provides a striking signature. The measurement of the leading protons allows to fully determine the kinematics of the central system, and well defined final states in the central detector matching the proton kinematics can then be selected and precisely reconstructed.

The exclusive two-photon production of pairs of photons, W and Z bosons provides a novel and unique testing ground for the electroweak gauge boson sector. The detection of 
$\gamma\gamma\rightarrow W^+W^-$ events allows to measure the quartic gauge coupling $WW\gamma\gamma$ with high precision. 
One can study the distributions and measure the production rates of these interactions, and verify whether they are compatible with the SM. An improved sensitivity of the order of $10^{-3}\div10^{-4}$ is expected with respect to earlier measurements~\cite{Chatrchyan:2013akv,Sirunyan:2017fvv,Aaboud:2017tcq}.
As a first step, the exclusive dilepton process $pp \rightarrow p\ell^+\ell^-p^{(*)}$ ($\ell=e,\mu$) 
has been observed for the first time at the LHC in pp collisions at $\sqrt{s}=13$~TeV~\cite{Cms:2018het}.
One of the two scattered protons is measured in the PPS detector, while the second proton either remains intact or is excited and then dissociates into a low-mass state $p^\star$, which is undetected.
Central dilepton production is dominated by the diagrams in which both protons radiate quasi-real photons that interact and produce the two leptons in a t-channel process.
The acceptance for detecting both protons in exclusive $pp\rightarrow p\ell^+\ell^-p$ events starts only above $m(\ell^+\ell^-)\approx$400~GeV, where the SM cross section is small. By selecting events with only a single tagged proton, the sample contains a mixture of lower mass exclusive and single-dissociation ($pp \rightarrow p\ell^+\ell^-p^*$) processes with higher cross sections.
A pair of leptons from a Drell-Yan process can also mimic a signal event if detected in combination with a pileup proton.
The kinematics of the dilepton system can be used to determine the momentum of the proton, and hence its fractional momentum loss $\xi$. Comparison of this indirect measurement of $\xi$ with the one obtained with PPS can be used to suppress backgrounds.
A total of $12~\mu^+\mu^-$ and $8~e^+e^-$ pairs with $m (\ell^+\ell^-) > 110$~GeV matching forward proton kinematics were observed, 
an excess of more than five standard deviations over the expected backgrounds.
The result constitutes the first observation of proton-tagged $\gamma\gamma$ collisions at the electroweak scale. 

At large $\sqrt{s_{\gamma\gamma}}$, the two-photon process $\gamma\gamma\rightarrow W^+W^-$ 
provides a window on BSM physics, since it is sensitive to triple and quartic gauge boson couplings. 
In pp collisions at $\sqrt{s}=8$~TeV, CMS has observed 13 candidate events in a final state with $e^\pm\mu^\mp$, large missing transverse energy, and no additional tracks, 
but without detecting the protons~\cite{Khachatryan:2016mud}. 
The observed yields and the kinematic distributions are compatible with the SM prediction for exclusive and quasi-exclusive $\gamma\gamma\rightarrow W^+W^-$ production.
The results are used to derive upper limits on the anomalous quartic gauge coupling (AQGC) parameters.
With an integrated luminosity of 100~fb$^{-1}$, the PPS is expected to improve the limits by at least two orders of magnitude, or perhaps observe a deviation from the SM production.
Among other interesting topics, the PPS can also probe the presence of composite Higgs and anomalous gauge-Higgs couplings, 
search for excited leptons, technicolor, extra-dimensions, axions, heavy exotic states, dark matter candidates, 
and explore more BSM processes~\cite{Delgado:2014jda,Inan:2010af,Lebiedowicz:2016lmn,Fichet:2013ola}.

\section{Summary}

The PPS extends the detector coverage to very forward regions and provides for additional sensitivity to BSM processes at the LHC. 
The detector has been regularly operating in high-luminosity fills and collected approximately 50~fb$^{-1}$ of data in 2016 and 2017. In 2018, it has collected a large fraction of the luminosity delivered. 
By tagging the leading outgoing protons from the primary interaction and correlating them with the events produced in the central detector, 
PPS can provide a clean selection of exclusive events that may reveal new and unexpected results, thus enhancing the physics potential of the CMS experiment.
First results of exclusive dilepton events have been obtained and provide for the first time the conditions to study particle production with masses at the electroweak scale via photon-photon fusion.

\ack

To my TOTEM and CMS colleagues who contributed to the success of this difficult project becoming a reality, without forgetting that there are many challenges ahead. 
To the Organizers for the kind invitation, and an interesting meeting in a secluded place.


\section*{References}
\begin{thebibliography}{9}
\bibitem{ctppstdr}
CMS and TOTEM Collaborations, ``CMS-TOTEM Precision Proton Spectrometer,'' 2014 CERN-LHCC-2014-021.

\bibitem{Antchev:2017pjj} 
  Antchev~G {\it et al.} [TOTEM Collaboration],
  ``Diamond detectors for the TOTEM timing upgrade,'' 2017
  {\it JINST} {\bf 12}, no. 03, P03007.

\bibitem{Berretti:2016sfj} 
  Berretti~M {\it et al.},
  ``Timing performance of a double layer diamond detector,'' 2017
  {\it JINST} {\bf 12}, no. 03, P03026.

\bibitem{Chatrchyan:2013akv} 
  Chatrchyan~S {\it et al.} [CMS Collaboration],
  ``Study of exclusive two-photon production of $W^+W^-$ in pp collisions at $\sqrt{s} = 7$~TeV and constraints on anomalous quartic gauge couplings,'' 2013
  {\it JHEP} {\bf 1307}, 116.

\bibitem{Sirunyan:2017fvv} 
  Sirunyan~A~M {\it et al.} [CMS Collaboration],
  ``Measurement of vector boson scattering and constraints on anomalous quartic couplings from events with four leptons and two jets in proton-proton collisions at $\sqrt{s}=$13~TeV,''
  2017 {\it Phys.\ Lett.\ B} {\bf 774}, 682.

\bibitem{Aaboud:2017tcq} 
  Aaboud~M {\it et al.} [ATLAS Collaboration],
  ``Study of $WW\gamma$ and $WZ\gamma$ production in pp collisions at $\sqrt{s} =$8~TeV and search for anomalous quartic gauge couplings,'' 2017
  {\it Eur.\ Phys.\ J.\ C} {\bf 77}, no. 9, 646.

\bibitem{Cms:2018het} 
  Sirunyan~A~M {\it et al.} [CMS and TOTEM Collaborations],
  ``Observation of proton-tagged, central (semi)exclusive production of high-mass lepton pairs in pp collisions at 13~TeV with the CMS-TOTEM precision proton spectrometer,'' 2018
  {\it JHEP} {\bf 1807}, 153.

\bibitem{Khachatryan:2016mud} 
  Khachatryan~V {\it et al.} [CMS Collaboration],
  ``Evidence for exclusive $\gamma\gamma \to W^+ W^-$ production and constraints on anomalous quartic gauge couplings in pp collisions at $ \sqrt{s}=7$ and 8~TeV,'' 2016
  {\it JHEP} {\bf 1608}, 119.

\bibitem{Delgado:2014jda} 
  Delgado~R~L, Dobado~A, Herrero~M~L and Sanz-Cillero~J~J,
  ``One-loop $\gamma\gamma \to$ W$_{L}^{+}$ W$_{L}^{-}$ and $\gamma\gamma \to$ Z$_{L}$ Z$_{L}$ from the electroweak chiral lagrangian with a light Higgs-like scalar,'' 2014
  {\it JHEP} {\bf 1407}, 149.

\bibitem{Inan:2010af} 
  Inan~S~C,
  ``Exclusive excited leptons search in two lepton final states at the CERN-LHC,'' 2010
  {\it Phys.~Rev.~D}~{\bf 81}, 115002.

\bibitem{Lebiedowicz:2016lmn} 
  Lebiedowicz~P, Luszczak~M, Pasechnik~R and Szczurek~A,
  ``Can the diphoton enhancement at 750 GeV be due to a neutral technipion?,'' 
  2016 {\it Phys.~Rev.~D}~{\bf 94}, no. 1, 015023.

\bibitem{Fichet:2013ola} 
  Fichet~S and Gersdorff~G~von,
  ``Anomalous gauge couplings from composite Higgs and warped extra dimensions,'' 
  2014 {\it JHEP} {\bf 1403}, 102.

\end{thebibliography}
\end{document}